# Experimental measurement methods and data on irradiation of functional design materials by helium ions in linear accelerator

R. A. Anokhin, V. N. Voyevodin, S. N. Dubnyuk, A. M. Egorov, B. V. Zaitsev, A. F. Kobets,
O. P. Ledenyov, K. V. Pavliy, V. V. Ruzhitsky, G. D. Tolstolutskaya

*Institute of Solid State Physics, Materials Science and Technologies*
*National Scientific Centre Kharkov Institute of Physics and Technology,*
*Academicheskaya 1, Kharkov 61108, Ukraine.*

The experimental research on the irradiation of the functional design materials by the *Helium* ions in the linear accelerator is conducted. The experimental measurements techniques and data on the irradiation of the functional design materials by the *Helium* ions with the energy up to *4 MeV,* including the detailed scheme of experimental measurements setup, are presented. The new design of accelerating structure of the *IH-type* such as *POS-4*, using the method of alternate-phase focusing with the step-by-step change of the synchronous phase along the focusing periods in a linear accelerator, is developed with the aim to irradiate the functional design materials by the *Helium* ions. The new design of the injector of the charged *Helium* ions with the energy of *120 KeV* at the output of an accelerating tube and the accelerating structure of the type of *POS-4* for the one time charged *Helium* ions acceleration in the linear accelerator are researched and developed. The special chamber for the irradiation of functional design materials by the *Helium* ions is also created. In the process of experiment, the temperature of a sample, the magnitude of current of *Helium* ions beam and the irradiation dose of sample are measured precisely. The experimental measurement setup and techniques are fully tested and optimized in the course of the research on the electro-physical properties of irradiated samples and the thermal-desorption of *Helium* ions in a wide range of temperatures.

## Introduction

At the present time, the generated energy at the nuclear power plants (*NPP*) constitutes more than *50 %* from the total energy, developed at the various power generation facilities in the *State of Ukraine*. However, in the nearest future, a considerable number of nuclear reactors at the *NPPs* in *Ukraine* is planned to be taken out of service, because of their operation time resource limitation. There is a similar tendency, as far as the old type nuclear reactors at the *NPP* is concerned, in many other countries. In this connection, the development of a new generation of nuclear reactors, including the fast neutrons nuclear reactor (*FNNR*) and the thermonuclear reactor (*TNR*), is considered. It is a well known fact that there would be a presence of the physical-chemical process of accumulation of considerable amount of *Helium*, which could be created as a result of the different nuclear reactions in the case of the *FNNR*; and also, it could be directly injected from the plasma in the case of the *TNR*. It is assumed that the *Helium* accumulation process will take place in the functional design materials of active zone in the *FNNR*, and especially, in the first wall in the *TNR*, along with a high degree of the radiation damages of functional design materials structures.

The *Helium* makes a strong influence on the radiation damageability of functional design materials, and it can frequently be a main reason of the considerable deterioration of physical properties of functional design materials, resulting in the service life reduction of designed elements in the nuclear reactors. In this connection, the scientists focus a serious attention to the research problem on the physical behaviour of *Helium* in relation to the various functional design materials at action by the different external conditions. A number of physical features on the *Helium's* influence in relation to the radiation damages of functional design materials, including the high- and low- temperatures radiation hardening and fragility, the radiation accelerated creep, the erosion of surfaces, in the first wall in the *TNR* and some other characteristics were researched [1 - 5]. Except the comprehensive knowledge on the influence by the different gases on the radiation damageability of functional design materials, the detailed experimental research data on the physical behaviour of different gases, depending on the various internal and external factors, are needed to be considered with the purpose to forecast the radiation damageability of functional design materials in the specified conditions of accumulation of considerable concentration of the gases. The following factors have to be taken to the consideration: the *Helium* penetration depth, the *Helium* spatial concentration in the distribution dependencies, the contents of impurities, defects and alloying elements in the metals and alloys with the various crystal grating systems, the processing of functional design materials after the irradiation process. At the present time, there is a considerable lack of obtained experimental research data on the accurate characterization of the physical behaviour of *Helium* in the functional design materials, depending on the above physical conditions.

At the research on the physical behaviour of *Helium* in the functional design materials, the special research attention is focused on the *Helium* influence on the mechanical properties of the functional design materials. It is a well-known fact that a presence of a few tens of *ppm* of *Helium* in the austenite steel results in the considerable fragility, which is accompanied by the change in the character of austenite steel destruction: from the trans-crystal cracking to the between-crystals cracking. In the case of the mechanical testing of samples with the known concentration of *Helium*, the special irradiating devices are needed. These irradiation devices can inject the *Helium* on the penetration depths from the tens of micrometers to the hundreds of micrometers into the sample; and they must have a dimension of the irradiation beam, without the scanning, up to several centimeters.

The main purpose of given research work is to conduct an experimental investigation on the irradiation of the newly synthesized functional design materials by the *Helium* ions in the linear accelerator with the energies from *0,12 MeV* to *4 MeV*, and the testing of this experimental technique at the research on both the *Helium* thermal desorption and the electrophysical properties of irradiated samples.

## 2. Experimental research techniques

The linear accelerator of the *Helium* ions with the energy of *4 MeV* was designed at the *National Scientific Centre Kharkov Institute of Physics and Technology* to make the advanced research on the *Helium* ions implantation into the samples [6]. The further increase of linear accelerator's energy up to the magnitude of *34 MeV* is planned [6].

### 2.1. Discussion on Helium ions accelerating structure of type of POS-4 and Helium ions injector

The accelerating section of *POS-4* is intended for the $He^+$ ions acceleration up to the energy of *4 MeV* (the initial energy is *30 KeV/nucleon*), moreover, it can be used for the acceleration of protons. In the accelerating *IH-type* structure of *POS-4*, the method of alternating phase focusing with the step-by-step change of the synchronous phase along the focusing periods is used to focus the *Helium* ions beam [7]. The efficiency of this method depends on the configuration of the focusing period. The choice of synchronous phases has provided an opportunity to capture the ions beam of high intensity in the phase angle of *120 °*; and also, to reach the radial and phase stabilities of the ions clots along the accelerating structure. The accelerating field had an increasing magnitude at the initial part of accelerating structure, aiming to provide the maximal capture of accelerated particles (*120 °*) in a mode of the stable radial and longitudinal movements. The scheme of accelerating structure of the type of *POS-4* and the amplitude of high frequency field, depending on the number of accelerating cells are shown in Fig. 1.

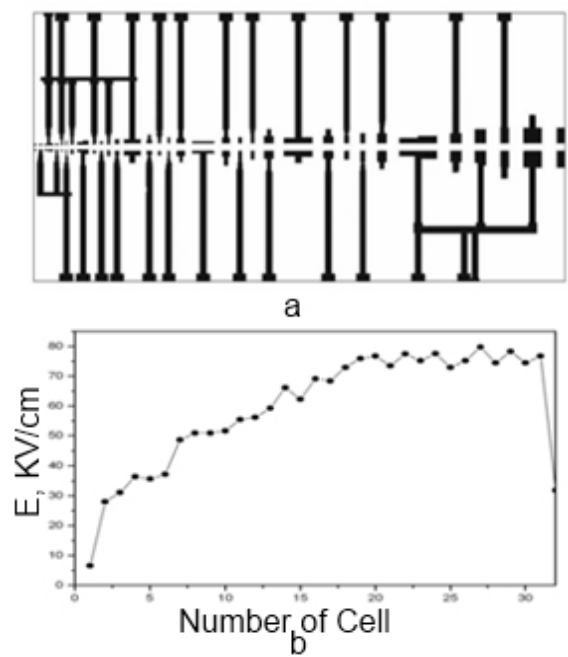

***Fig. 1.*** *Scheme of accelerating structure of type of POS-4 (a) and the amplitude of high frequency field, depending on the number of accelerating cells (b).*

In the result of both the computing modeling and the experimental measurements, which were completed in the processes of testing and tuning of the accelerating structure of the type of *POS-4*, the data on the distribution of electric field in the accelerating structure at the operation frequency of *47,2 MHz* is obtained. The main parameters of the accelerating structure of the type of *POS-4* are presented in Tab. 1.

The injector of the charged *Helium* ions with the energy of *120 KeV* for the injection of *Helium* ions beam into the accelerating structure of the type of *POS-4* is developed with the purpose to irradiate the functional design materials by the *Helium* ions. The injector consists a source of the *Helium* ions, the beam focusing system, and also, the accelerating tube. The double-plasmatron source with the oscillating electrons in the anode region was used to generate the *Helium* ions beam of specified parameters [8].

| Operating frequency, *MHz* | 47,2 |
|---|---|
| Input impulse current, *mA* | 30 |
| Output impulse current, *mA* | 1…0,3 |
| Output normalized *rms* emittance in plane X, *cm·mrad* | 0,0584 |
| Output normalized *rms* emittance in plane Y, *cm·mrad* | 0,0568 |
| Output normalized 99% emittance in plane X, *cm·mrad* | 0,7809 |
| Output normalized 99% emittance in plane Y, *cm·mrad* | 0,7723 |

***Tab. 1.** Main technical parameters of accelerating structure of type of POS-4.*

Such choice has been made, because the *Helium*, which has an abnormal *Pashen* curve, is the working gas. In the source, the arc discharge is created to obtain the plasma, which penetrates from the anode region to the slot-hole between the plasma part of a source and the extracting electrode, where the double layer, which is a source of the *Helium* ions, is formed at the application of extracting potential.

The injector allows to generate the beam of one time charged *Helium* ions with the current magnitudes of a few tens of milliampere. The main parameters of injector are presented in Tab. 2.

| Working gas | Helium |
|---|---|
| Current of arc discharge, *A* | 2…4 |
| Current of beam at output, *mA* | up to 20 |
| Energy of particles at output, *KeV* | 120 |
| Diameter of beam at output, *mm* | ~ 8 |
| Pressure of working gas in anode region of source, *mm of mercury column* | $5·10^{-3}$ |
| Transmission frequency, *Hz* | 2…10 |
| Duration of impulse of discharge arc modulator, *μsec* | 500 |
| Magnetic field magnitude in source, *Oe* | 300…700 |

***Tab. 2.** Main technical parameters of injector.*

In Fig. 2, the oscillogram of dependence of the accelerated *Helium* ions beam current on the time is presented. The arch discharge current is equal to *4 A*, the injection current is *1,2 mA*, the current of the accelerated *Helium* ions beam is ~ *300 μA*. It can be seen that the accelerated *Helium* ions beam current has a quasi-squared shape, the deviation of the peak magnitude of current is not more than *5 %*.

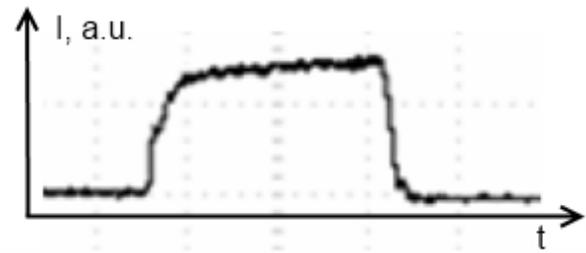

***Fig. 2.** Oscillogram of Helium ions beam current at output of accelerating structure of type of POS-4 with energy of 4 MeV.*

### 2.2. Discussion on target section in linear accelerator and experimental conditions of samples irradiation

The special chamber within the target section has been developed to irradiate the samples of functional design materials by the *Helium* ions. The scheme of target section is shown in Fig. 3. The temperature of a sample in the irradiation process was measured by the *Chromel – Alumel* thermocouple attached to the sample from an opposite side in relation to the incoming *Helium* ions beam (see Fig. 3). The signal from the thermocouple was amplified by the differential amplifier. The calibration of thermocouple was made by taking into an account the length of measuring wires (~ *30 m*).

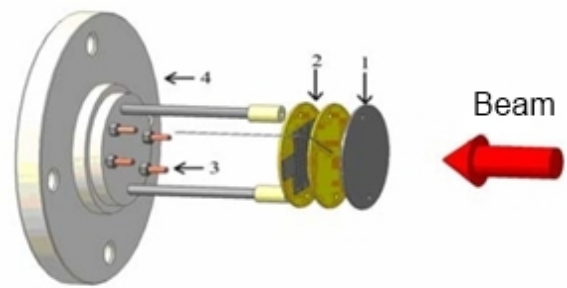

***Fig. 3.** Scheme of target section: 1 – sample; 2 – heater; 3 – current inputs; 4 – flange.*

The measurements of the magnitudes of currents of the *Helium* ion beams at the input and output ports of a linear accelerator were made by the induction sensors. During the experimental research, the currents of both the injected *Helium* ions beam with the energy of *120 KeV* and the accelerated *Helium* ions beam with the energy of *4 MeV* were measured. The current of the *Helium* ions beam was measured by the induction sensors, mounted in front of the sample on the distance of ~ *30 cm*. Before the each next irradiation process, the calibration of induction sensors was made, using the *Faraday* cylinder. The average diameter of *Helium* ions beam, falling on the sample, was ~ *36 … 37 mm*; and the area of the irradiated sample was around *10 $cm^2$*. The number of samples, which were irradiated by the *Helium* ions in the linear accelerator was 10. The irradiated functional design materials and the parameters of irradiation are presented in *Tab. 3*.

| Number of sample | Sample | Thickness of sample, $\mu m$ | Energy of He$^+$ ions, MeV | Mean temperature of irradiation, °C | Complete irradiation dose, part./cm$^2$ |
|---|---|---|---|---|---|
| 1 | X18H10T | 200 | 4 | 72 | $7.5 \cdot 10^{15}$ |
| 2 | Steel-3 | 100 | 4 | 78 | $1.5 \cdot 10^{16}$ |
| 3 | Zr+2.5%Nb | 250 | 4 | 70 | $2.3 \cdot 10^{16}$ |
| 4 | Zr | 300 | 4 | 58 | $5 \cdot 10^{16}$ |
| 5 | Zr+1%Nb | 250 | 2.42 | 40-80 | $5 \cdot 10^{16}$ |
| 6 | Zr+1%Nb | 250 | (4+0.12) | 40-80 | $5 \cdot 10^{16} + 5 \cdot 10^{16}$ |
| 7 | Zr+1%Nb | 250 | 0.12 | 40-80 | $5 \cdot 10^{16}$ |
| 8 | Zr+1%Nb | 250 | 0.12 | 40-80 | $5 \cdot 10^{17}$ |
| 9 | Nb | 16 | 4 | 40-80 | $5 \cdot 10^{16}$ |
| 10 | Nb+1%Zr | 22 | 4 | 40 | $5 \cdot 10^{16}$ |

***Tab. 3.*** *Irradiated materials and parameters of irradiation.*

## 2.3. Discussion on measurement techniques

During the irradiation process, the measurements of the sample's temperature at the irradiation, the current of *Helium* ions beam, the irradiation dose were conducted. The analogue-digital converter of the model of *ZET 210 «Sigma USB»*, connected to the computer, was used. The software program was written with the purpose to automate the precise measurements and to save the experimental data. The measurement setup scheme for the measurements of the sample's temperature at irradiation, the current of *Helium* ions beam, the irradiation dose are shown in Fig. 4.

The sample's temperature is measured with the help of the *Chromel – Alumel* thermocouples with the frequency of *1 Hz*, allowing to detect it's change at the irradiation precisely. The thermocouple is connected to the input of the analogue to digital converter (*ADC*), which transfers the digitized data on the thermal electromotive force to the constant voltage voltmeter. After the voltmeter, the measured data is shown on the display in the form of the graph with the help of the *XYZ*-plotter, and also, on the digital indicator (the data at the present time moment), and is saved as a file in the hard disk of computer for the subsequent analysis.

The measurement of *Helium* ions beam current density is made, using the following scheme. After the analogue data digitization by the *ADC*, the digital data is transferred to the alternating voltage voltmeter, which is adjusted for the measurement of the peak voltage magnitude. This magnitude is shown on the screen in the form of the graph with the help of the *XYZ*-plotter) and on the digital indicator (magnitude at the present time moment). The measured data are saved in a file on a hard disk in the computer. The measurements are made at the frequency of *10 Hz*, allowing to detect the peak magnitude of *Helium* beam current most precisely.

The measurement of irradiation dose of a sample is conducted, using the scheme similar to the scheme of measurement of the peak magnitude of the *Helium* ions beam current. The only distinction is that the integrator is placed after the voltmeter. It allows the direct registering of the irradiation dose of a sample during the experimental works at the linear accelerator.

In Fig. 5, the front panel of the *Helium* ions beam current measurement device is shown. The temperature measurement device and the irradiation dose measurement device have similar front panels.

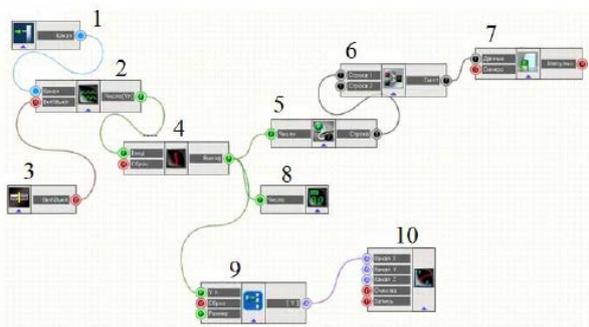

***Fig. 4.*** *Scheme of measurements of 1) sample temperature at irradiation, 2) current of Helium ions beam, 3) irradiation dose: 1 – input channel; 2 – voltmeter, 3 – on/off switch; 4 – integrator, which is only used for measurement of irradiation dose; 5 - ADC converter; 6 – summation of data; 7 – saving of data in file; 8 – liquid crystal indicator; 9 – data massive; 10 – XYZ plotter.*

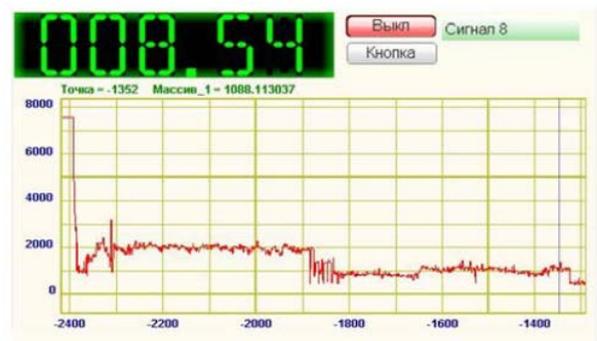

***Fig. 5.*** *Display of measurement device for precise characterization of Helium ions beam current density.*

## 2.4. Discussion on methodical experiments

In the course of the sample's irradiation experiment, the various operation modes of experimental setup, including both the linear accelerator and the registration devices, were tested firstly. The influence by the *on/off* switching of power supply of the high frequency accelerating structure block was also tested. The sensitivity is in the limits of *0,5 °C* as it follows from the graph in Fig. 6. The temperature of a sample is stabilized after the linear accelerator operation time of *60 ... 90 min*.

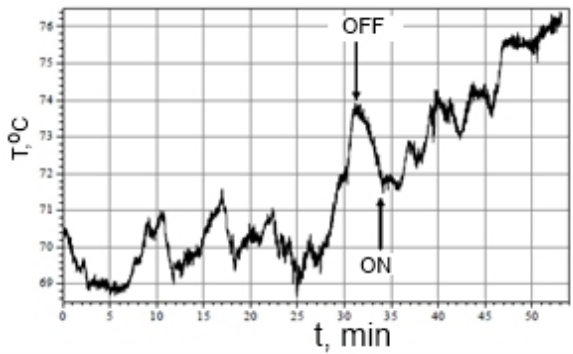

***Fig. 6.*** *Dependence of temperature of X18H10T sample on time of irradiation.*

The measurements of the *Helium* ions beam current (Fig. 7) have shown that the current of the accelerated *Helium* ions up to *4 MeV* is in *4-6* times smaller than the current of the injected *Helium* ions. The stability of the *Helium* ions beam current at the optimal regime of the linear accelerator operation is ± *5 %*, and it is also defined by the stability of the power supply systems.

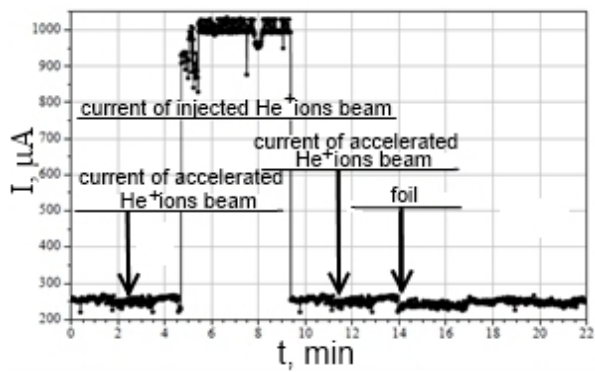

***Fig. 7.*** *Dependence of peak magnitude of Helium ions beam current on time of irradiation*

The irradiation dose is registered with the help of the induction sensors of the *Helium* ions beam current and the programmed integrator. The dependence of an irradiation dose of a sample on the time is shown in Fig. 8. The conducted calculations confirm a right experimental techniques approach, which was used to register the irradiation dose of a sample.

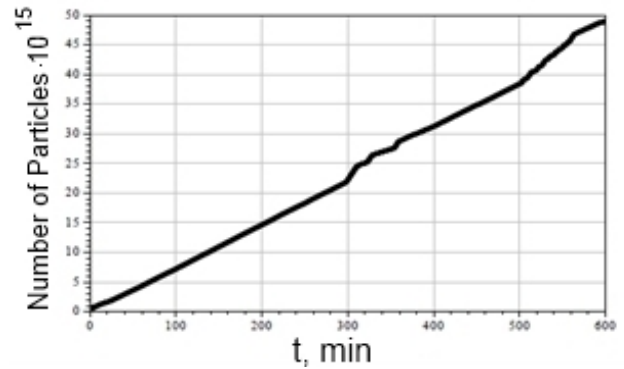

***Fig. 8.*** *Dependence of magnitude of irradiation dose of Zirconium sample on time.*

During the completion of methodical experiments, the great attention has been paid to the problem of the *Helium* ions energy detection in the case of the falling *Helium* ions on the sample. The aluminum foil with the thickness of *10 μm* was placed in the front of an induction sensor with the purpose to control the energy of the *Helium* ions. The measurements of the *Helium* ions beam current without the foil and with the foil have resulted in the almost identical magnitudes of current (see Fig. 7). It is possible to state that the *Helium* ions with the energy of *4 MeV* reach the sample, because the *Helium* ions with the energy of *4 MeV* with the tracking paths of more than *16 μm* can penetrate the *Aluminum* foil freely.

In Fig. 9, the dependences of the sample's temperature (the top curve) and the *Helium* ions beam current (the bottom curve) on the time at the on/off switching of the power supply of the linear accelerator are shown. In this case, the energy of the injected *Helium* ions is *120 KeV*, and the energy of the accelerated *Helium* ions is *4 MeV*. The injected *Helium* ions beam current is in the *4* times bigger than the accelerated *Helium* ions current, and the temperature of a sample at the injected *Helium* ions beam current is smaller, than at the accelerated *Helium* ions beam current. Thus, an increase of the temperature of a sample is stipulated by the high-energy component of the *Helium* ions beam (*4 MeV*). The given conclusion completely coincides with the theoretical calculations and dependences, which are shown in Figs. 7, 9.

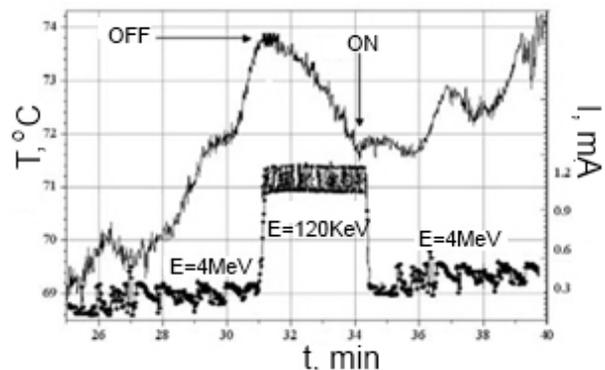

***Fig. 9.*** *Dependences of temperature of sample and Helium ions beam current on time at injected and accelerated Helium ions beams currents.*

As it follows from the conducted methodical experiments, pursuing the goal to stabilize and increase the *Helium* ions beam current, it is necessary: 1) to control the change of the high frequency field of the accelerating structure during the sample's irradiation process, and also, 2) to measure the distribution of the *Helium* ions current at the target for the purpose of maintenance of its uniformity in the region of the sample. It is also necessary to make the additional vacuum pumping in the position of sample with the purpose to reach the high vacuum in the chamber.

## 3. Experimental measurement results

In the case of irradiated sample, the computer modeling on the *Helium* distribution along the thickness of a sample were conducted with the use of the *SRIM* software. In Fig. 10, the dependences in the researched case of *Zr+1%Nb* are shown. The full irradiation doses of samples by the *Helium* ions with the energies of *120 KeV* and *4 MeV* are $5 \cdot 10^{16}$ *ions/cm²*. The measurements on the thermal desorption of *Helium* were completed, using the samples no. *1-8* (see Tab. 3).

The electrophysical properties were precisely characterized, using the samples of *Nb* and *Nb+1%Zr* (see Tab. 3). The *Helium* ions with the energy of *4 MeV* penetrate into the researched metals on the depths of *7,2* and *7,4* microns correspondingly. The main peaks of distribution of the implanted *Helium* ions are observed at these depths in the irradiated metal samples.

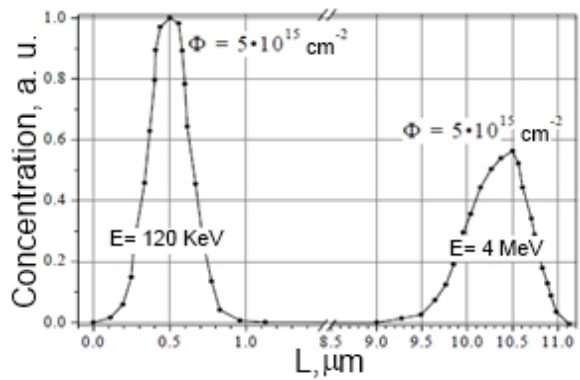

*Fig. 10. Computed distribution of Helium ions, which are implanted into Zr+1%Nb with energies of 120 KeV (the left graph) and 4 MeV (the right graph); irradiation dose is $5 \cdot 10^{15}$ 1/cm².*

### 3.1. Helium thermo-desorption from researched samples

The physical behaviour of *Helium* in the investigated materials after their irradiation by the *Helium* ions with energy of *0,12 ... 4 MeV* are studied, applying the thermo - desorption (*TD*) method. The *TD* technique includes the measurement of partial pressure of *Helium*, which outgoes from an investigated sample, in the process of post-implantation linear increase of its temperatures. In the experiments, the technique of thermo-desorption in the dynamic mode of operation at which the pressure of gas in the chamber is proportional to the velocity of desorption from the metal. The samples were investigated in the temperatures interval of *0 ... 1500 °C*, the velocity of their heating was *5 ... 8 °C/sec*.

The *TD* researches are completed on the experimental setup «Ant» [10] with the application of pumping system without the oil, which provides the pressure of residual gases in the target's chamber at the level of $(2 ... 3) \cdot 10^{-5}$ *Pa*. The structural analysis of the gas medium in the experimental chamber of measurement setup was conducted by the mass - spectrometer.

In Fig. 11-13, the *Helium* thermo-desorption spectrums from the steel of the type of *X18H10T* and from the alloy of *Zr+1%Nb* at the various sample's irradiation parameters are presented.

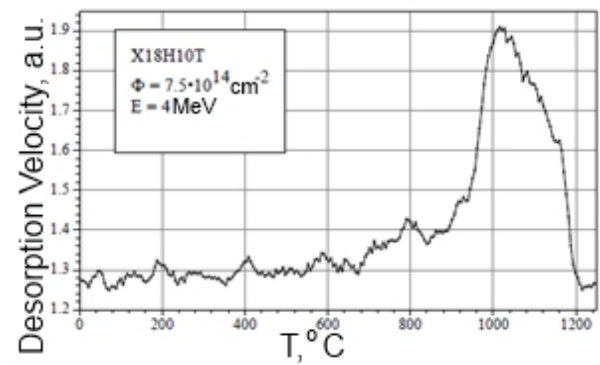

*Fig. 11. Helium thermo-desorption spectrum in sample of steel of type of X18H10T, which is irradiated by He⁺ ions with energy of 4 MeV. Heating velocity of sample is 5,5 °C/sec.*

There is a narrow range of temperatures with no desorption process in the case of the alloy of *Zr+1%Nb* at the certain magnitude of the *Helium* ions irradiation doses and without the dependence on the irradiation energy. There is the notable *Helium* gas outgo at $T \approx 500\,C$ and it proceeds up to $T \approx 1500\,°C$. In this range of temperatures, the *TD* spectrum is characterized by the superposition of several peaks of desorption. The complexity of structure of a spectrum increases at an increase of the irradiation dose (see Figs. 12, 13). The maximum gas outgo is observed in the peak with the temperature of maximum $T \approx 1280\,°C$. It makes sense to note that this stage of desorption is observed in the spectrums of samples, which are irradiated up to the dose $\Phi = 5 \cdot 10^{15}$ *cm⁻²* as well as in the samples, which are irradiated to the dose in *10* times more bigger.

In Fig. 13, the *TD* spectrums of the samples of *Zr+1%Nb* alloy, which are irradiated by the *Helium* ions with the energy of *2,42 MeV* up to the dose $\Phi = 5 \cdot 10^{15}$ *cm⁻²*, and also, the *TD* spectrums of the samples, which are consistently irradiated by the *Helium* ions with the energies of *0,12 MeV* and *4 MeV* up to the same irradiation dose of $\Phi = 5 \cdot 10^{15}$ *cm⁻²*. Considering the *TD* spectrums, which are shown in Fig. 12, it is necessary to note that the first spectrum differs by the increased value of the *Helium* gas atoms at the main high temperature peak.

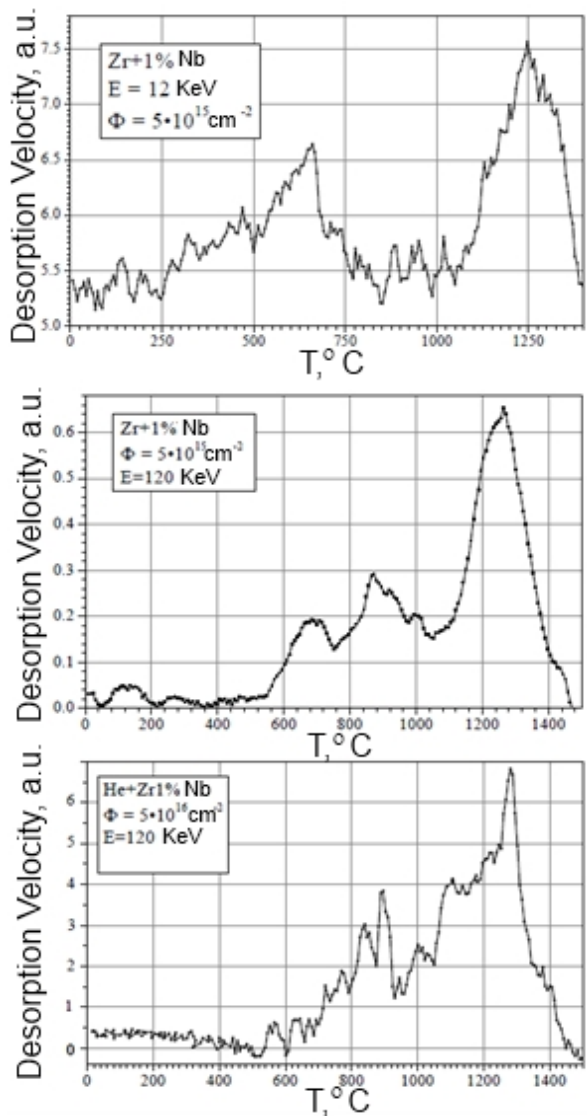

***Fig. 12.** Helium thermo-desorption spectrum in sample of Zr+1%Nb, which is irradiated by He⁺ ions with energies of 12 and 120 KeV. Heating velocities of samples: 5 °C/sec – 12 KeV; 7,8 °C/sec – 120 KeV.*

The research on the *Helium* gas outgo process from the *Zr+1%Nb* samples after their consecutive irradiation by the *Helium* ions has shown that the structure of spectrums in this case is less complicated, comparing to the spectrums, which are obtained at the irradiation of the *Zr+1%Nb* alloy by the *He⁺* ions with the energies of *120 KeV* and *2,42 MeV*. The notable desorption begins at the temperature $T \approx 600$ °C. The peak of the *Helium* gas outgo with a maximum at the temperature $T \approx 1200$ °C prevails in the spectrum. At the irradiation of samples of the *Zr+1%Nb* alloy by the *Helium* ions with the energies of *0,12 MeV* and *4 MeV* up to the same irradiation dose $\Phi = 5 \cdot 10^{15}$ cm$^{-2}$, the spectrum of thermo-desorption changes slightly. As it is visible in Fig. 13, in this case, the spectrum is also less complicated in comparison with the spectrums, which are obtained at the irradiation of the *Zr+1%Nb* alloy by the *He⁺* ions with the energies of *120 KeV* and *2,42 MeV*. The additional experiments are necessary to precisely characterize the influence by the penetration depths of the implanted particles and the created radiation damages by the implanted particles in the crystal gratings on the mechanisms of the *Helium* outgo, and hence the structure of the *TD* spectrum in the volume of sample.

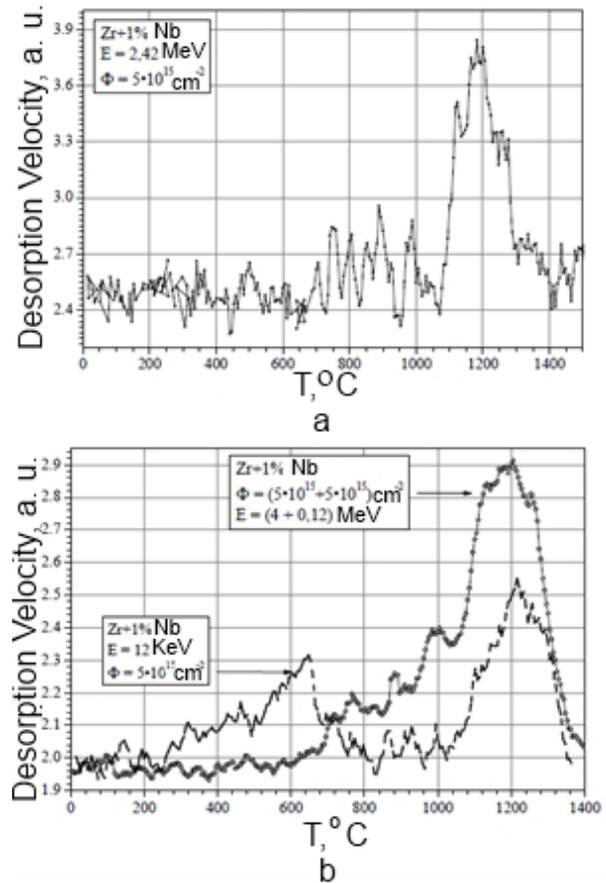

***Fig. 13.** Helium thermo-desorption spectrum in sample of Zr+1%Nb, which was irradiated by Helium ions (He⁺) ions with energies of:*
*a) 2,42 MeV;*
*b) 4 MeV+120 KeV (top curve), and*
*120 KeV (bottom curve).*
*Heating velocity of samples: 5,6 °C/sec – 2,42 MeV;*
*8 °C/sec – 4 +0,12 MeV; 5 °C/sec – 12 KeV.*

An observable absence (or a very small intensity) of the *Helium* outgo from the steel as well as from the alloy of *Zirconium* and *Niobium* at the temperatures much below $T \approx 500 ... 650$ °C confirms that the contents of the *Helium* atoms in the *Zr+1%Nb* alloy does not change at the heating in this range of temperatures. Probably, in this range of temperature, there are the complex diffusion processes of the *Helium* subsystem and the defective structure of a crystal lattice transformations, which result in the multistage processes of the *Helium* outgo from the implanted *Helium* volume at the much high temperatures. The presence of several peaks in the *TD* spectrums points out to an existence of several discrete stages of the *Helium* outgo with the different thermodynamic characteristics and the *Helium* outgo mechanisms from the metal. At the irradiation process, the *Helium* atoms interact with the defects of crystal grating, which can be considered as the traps for the *Helium*. During this process, there are the *Helium* capture by the single vacancies, double-vacancies, and also, congestions of vacancies (the strong traps for the *He* atoms); the *Helium* capture by the dislocations and boundaries of grains, the inter-phase boundaries; and the formation of *Helium* and *Helium-vacancy* clusters [3]. In Fig. 13, the *TD* spectrum of *Helium* in the *Zr+1%Nb* alloy is shown by the dashed line. The *Helium* is implanted by the continuous *Helium* ions beam with the energy of *12 KeV* up to the irradiation dose $\Phi = 5 \cdot 10^{15} cm^{-2}$. Going from the comparison of spectrums, it can be seen that, despite the considerable divergences in the depths of the implanted *Helium* atoms, the *Helium* outgo takes place in the peaks with the close temperatures in the high-temperature regions. It allows to make a conclusion that the same physical mechanism is responsible for the process. In the spectrum of samples with the implanted *Helium* ions with the low energy, there is the low temperature stage of the *Helium* desorption. Possibly, it is connected with the *Helium* outgo process from the *Helium – vacancy* systems, which are created in the sub-surface regions of a sample, and have the much high concentration, comparing to the case of the high energy irradiation.

The experimental fact that the *Helium TD* spectrums in the *Zirconium-Niobium* alloy are close enough in the range of high temperatures at the various *Helium ions* saturation mechanisms in the samples apparently has an important meaning at the consideration of the problem about the adequacy of both the technique of impulse implantation of materials and the technique of continuous irradiation (see, for example, [11]).

In our opinion, the innovative research on the *Helium* atoms *TD* nature in the functional design materials at the irradiation conditions as well as the research on the existing connections between the *Helium* atoms *TD* process and the migration of separate atoms or their systems to the surface, which can be facilitated by the created radiation defects and by the micro holes of gas, have to be continued.

### 3.2. Discussion on electrophysical properties of irradiated functional design materials

The irradiation of metal samples, which were made of the thin foils of the high pure *Nb* and the *Nb+1 % at. Zr* alloy (*see no. 9, 10 in Tab. 3*) by the $He^+$ ions with the energies of up to *4 MeV* and the flounce of $5 \cdot 10^{15}$ *ions/cm²* at the temperature of *40 °C* was completed. The thicknesses of samples are *16* and *22 microns* correspondingly. The samples were simultaneously exposed to the irradiation during the one experimental stage, and they had the same values of flounces. The mean magnitudes of the specific resistances *ρ* of the samples were measured before and after the irradiation process at the room and nitrogen (*-196 °C*) temperatures with the application of the measuring scheme with the four sensors, using the potentiometers of the model of *P363-1*. After the irradiation, the specific resistance of *Nb*-sample at the room temperature was on *4,18 %* bigger than before the irradiation, whereas in the case of the sample of *Nb+1%Zr*, the irradiation influence was on an order of magnitude smaller. At the nitrogen temperature, in view of the decrease of the electrons scattering on the phonons, the contribution by the implanted *Helium* ions in the magnitude of specific resistance appeared to be a little bit bigger and was equal to *5,47 %*. The weak influence by the introduced *Helium* atoms on the magnitude of specific resistance in the sample of *Nb+1%Zr* alloy follows from the fact that a big enough quantity (*∼ 1 at. %*) of *Zr* impurities is introduced into the alloy. The *Zr* atoms represent the main scattering centers for the electrons, because they have the significantly bigger dimensions, comparing to the small sizes of the *He* atoms. Considering the non-homogeneity in the distribution of the introduced *Helium* atoms along the thickness of a sample, the magnitudes of electrical specific resistance in the samples represent the averaged magnitudes only, whereas in the region of the main accumulation of *Helium* in the core of a sample, the changes of electrical specific resistance can probably be much bigger. The technique of the *Helium* homogeneous implantation along the thickness of a sample is supposed to be developed during our next stage in the research on the electrophysical properties of irradiated metal samples.

### Conclusion

The experimental research on the irradiation of the nuclear reactor functional design materials by the *Helium* ions in the linear accelerator is completed. The experimental measurements methods and data on the irradiation of the nuclear reactor functional design materials by the *Helium* ions with the energy up to *4 MeV,* including the detailed scheme of experimental measurements setup, are presented. The new design of an accelerating structure of the *IH-type* such as *POS-4*, using the method of alternate-phase focusing with the step-by-step change of the synchronous phase along the focusing periods in a linear accelerator, is developed with the aim to irradiate the nuclear reactor functional design materials by the *Helium* ions. The new design of the injector of the charged *Helium* ions with the energy of *120 KeV* at the output of an accelerating tube and the accelerating structure section of the type of *POS-4* for the one time charged *Helium* ions acceleration in the linear accelerator are researched and developed. The special chamber for the irradiation of the nuclear reactor functional design materials by the *Helium* ions is also created. In the process of experiment, the temperature of a sample, the magnitude of current of the *Helium* ions beam and the irradiation dose are measured precisely. The experimental measurement setup and techniques are fully tested and optimized in the course of the research on the electro-physical properties of irradiated samples and the thermal-desorption of *Helium* ions in a wide range of temperatures.

Authors are very grateful to a group of distinguished scientists at the *National Academy of Sciences in Ukraine (NASU)*, led by *Boris E. Paton*, for the numerous encouraging scientific discussions on the reported experimental research results on the irradiation of the nuclear reactor's functional design materials by the *Helium* ions in the linear accelerator.

This innovative research is completed in the frames of the nuclear science and technology fundamental research program, facilitating the synthesis of new functional design materials for the fast neutrons nuclear reactors and the thermonuclear reactors at the *National Scientific Centre Kharkov Institute of Physics and Technology (NSC KIPT)* in Kharkov in Ukraine.

The research is funded by the *National Academy of Sciences in Ukraine (NASU)*.

This research article was published in the *Problems of Atomic Science and Technology* (*VANT*) in [12] in 2012.

[*]E-mail: ledenyov@kipt.kharkov.ua

———————